\documentclass{article}

\usepackage{arxiv}

\usepackage[utf8]{inputenc} % allow utf-8 input
\usepackage{textcomp}
\usepackage{txfonts}
\usepackage{hyperref}       % hyperlinks
\usepackage{url}            % simple URL typesetting
\usepackage{booktabs}       % professional-quality tables
\usepackage{amsfonts}       % blackboard math symbols
\usepackage{nicefrac}       % compact symbols for 1/2, etc.
\usepackage{microtype}      % microtypography
\usepackage{lipsum}
\usepackage{graphicx}
\usepackage{cite} 
\usepackage{xcolor}

\title{Core but not Peripheral Online Social Ties is a Protective Factor against Depression: Evidence from a Nationally Representative Sample of Young Adults}

\author{
  Sofia Dokuka \\
  Institute of Education\\
  National Research University\\ Higher School of Economics\\
  Moscow 101000, Russia \\
  \texttt{sdokuka@hse.ru} 
   \And
  Elizaveta Sivak \\
  Institute of Education\\
  National Research University\\ Higher School of Economics\\
  Moscow 101000, Russia \\
  \texttt{esivak@hse.ru} 
   \And
  Ivan Smirnov \\
  Institute of Education\\
  National Research University\\ Higher School of Economics\\
  Moscow 101000, Russia \\
  \texttt{ibsmirnov@hse.ru} 
}

\begin{document}
\maketitle

\begin{abstract}
As social interactions are increasingly taking place in the digital environment, online friendship and its effects on various life outcomes from health to happiness attract growing research attention. In most studies, online ties are treated as representing a single type of relationship. However, our online friendship networks are not homogeneous and could include close connections, e.g. a partner, as well as people we have never met in person. In this paper, we investigate the potentially differential effects of online friendship ties on mental health. Using data from a Russian panel study ($N = 4,400$), we find that – consistently with previous research – the number of online friends correlates with depression symptoms. However, this is true only for networks that do not exceed Dunbar’s number in size ($N \leq 150$) and only for core but not peripheral nodes of a friendship network. The findings suggest that online friendship could encode different types of social relationships that should be treated separately while investigating the association between online social integration and life outcomes, in particular well-being or mental health. 
\end{abstract}

% keywords can be removed

\section*{Introduction}
\noindent Individual’s mental health is known to be associated with their position within the social network. One of the most well-established relationships is the association between social integration and depressive symptoms \cite{kawachi2001social, van2010course, taylor2018social, elmer2020depressive, choi2020exposure}. Generally, social integration is understood as a structural aspect of people’s relationships, that indicates how those relationships are patterned or organized \cite{thoits2011mechanisms}. However, in most of the studies, social integration is simply defined as the number of social contacts.

\subsection*{Social integration and depression}

Longitudinal studies demonstrate that the association between social integration and depression might be explained by the protective role of social connections that serve as a stress buffer mitigating the depressive symptoms \cite{fiore1983social, hays2001does, thoits2011mechanisms}, or by the changes in friendship formation and interactions of individuals with depressive symptoms (i.e. such individuals tend to withdraw from existing contacts or create fewer new connections) \cite{coyne1976depression, negriff2019depressive}. Based on the National Longitudinal Study of Adolescent Health data (Add Health, $N = 11,023$), Ueno studied the association between the depressive symptoms and a variety of ego-network patterns \cite{ueno2005effects}. He concluded that the number of friends was the strongest predictor of depressive symptoms. Although other variables generally showed significant correlations with depressive symptoms in the expected directions, the associations were very weak, especially when controlling for the number of friends. Employing the same longitudinal dataset Shaefer et al. \cite{schaefer2011misery} analyzed the role of depressive symptoms in the evolution of friendship networks and demonstrated that depressed persons withdrew from friendships over time, leaving them with fewer friends. Depressed individuals were also less likely to be selected as friends by others because they tend to occupy peripheral network positions. Negriff \cite{negriff2019depressive} found that higher levels of depressive symptoms led to smaller, less connected networks with fewer friends in the largest connected component of the network two years later. She concluded that individuals with depressive symptoms lack the social skills needed to form and maintain close relationships, which leads to the dissolution of friendship ties.

Previously reported results on the relationship between social network structure and depressive symptoms were mostly obtained for complete networks based on self-reported data \cite{schaefer2011misery, elmer2020depressive}, e.g. by asking school students to nominate their friends. Having information about the whole network structure allows controlling for a variety of network effects. However, it limits the generalizability of the findings as it is not clear if they are specific to particular schools or could be generalized to a large population. As a result, the association between depressive symptoms and ego network structure is not well understood at the population level. Self-reports on social networks can also be biased \cite{latkin2017relationship} which makes it essential to search for more objective measures of social interactions.

\subsection*{Online social integration}

In the past decade, a significant fraction of social interactions migrated online, for example to social media platforms \cite{wiederhold2020connecting, liu2020online}. This process was accelerated in 2020 due to the COVID-19 pandemic and associated restrictions on face-to-face meetings, making social media and other digital platforms one of the key communication tools for a large part of the population. Given the central role that social media plays in interpersonal communication, it is important to understand the relationship between the online social environment of an individual and their depressive symptoms.

Empirical results for online networks largely agree with studies of offline networks. Users suffering from depression have smaller networks \cite{vedula2017emotional, park2013activities, negriff2019depressive} with densely clustered pockets and less frequently explicitly mention their network partners compared to the non-depressed users \cite{vedula2017emotional}. Individuals with suicide ideation and depressive symptoms also have less clustered personal social networks and tend to connect with individuals similarly oriented toward suicide and depression \cite{masuda2013suicide}.

Existing literature on the association between online networks and depressive symptoms mostly does not differentiate online friends, although online networks are known to consist of multiple layers, which differ in both emotional closeness with alters \cite{dunbar2015structure} and levels of support \cite{choi2020ten}. Moreover, network studies based on data on offline networks highlight the distinct impact of different ties on personal well-being \cite{granovetter1973strength, thoits2011mechanisms, elmer2017co}. This can also apply to online networks. For instance, Lup et al. \cite{lup2015instagram} showed that the association between Instagram use and depressive symptoms is moderated by the number of strangers followed: more Instagram use is related to greater depressive symptoms only for those at highest levels of strangers followed, and for those at lower levels, Instagram use and depressive symptoms are unrelated.

In this paper, we investigate the potentially differential effect of online friendship ties on mental health, based on the survey data and data from a social networking site in a sample of young adults.

\begin{figure}[t]
\centering
\includegraphics[width=0.5\columnwidth]{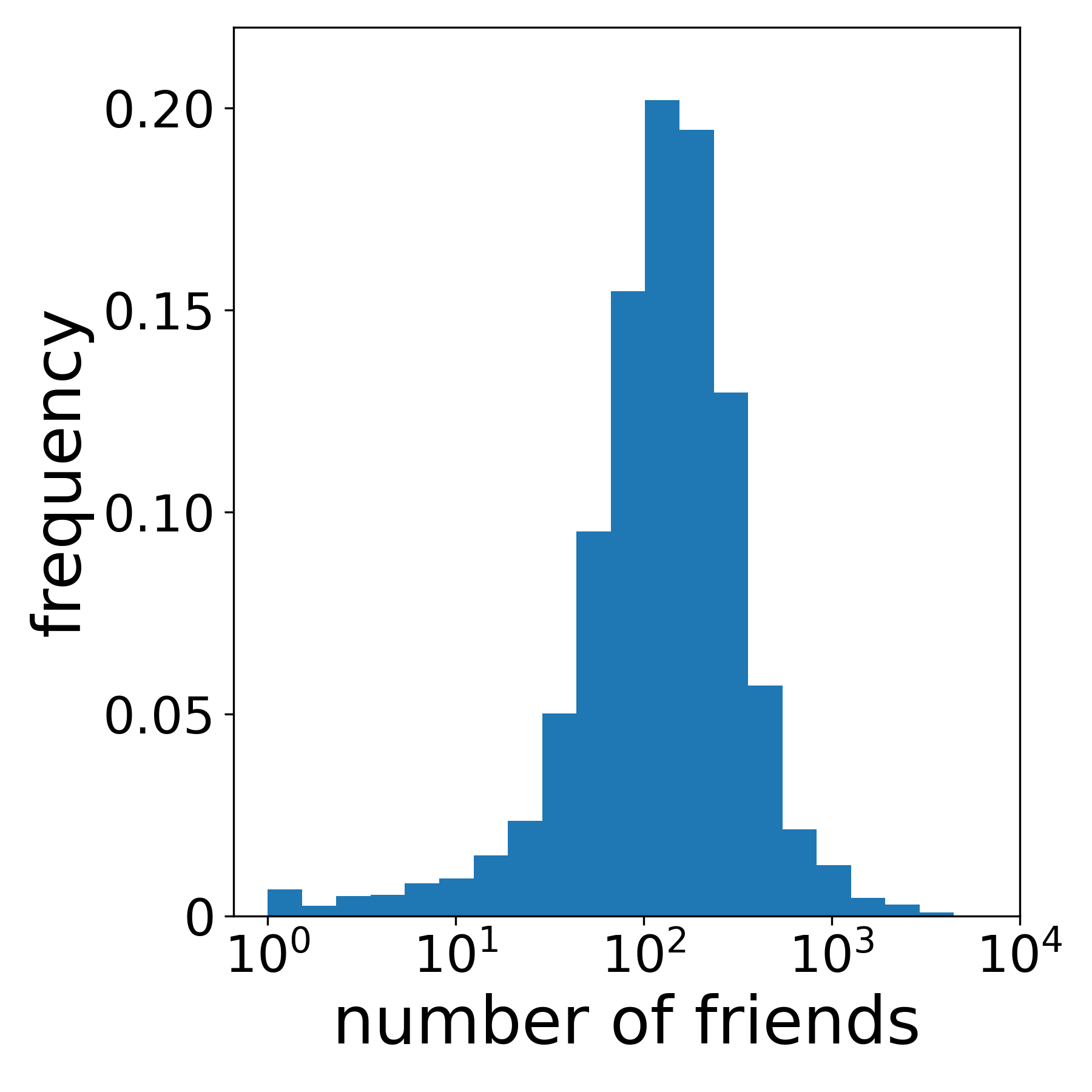} % Reduce the figure size so that it is slightly narrower than the column. Don't use precise values for figure width.This setup will avoid overfull boxes.
\caption{\textbf{The distribution of the number of online friends}. The median number of friends is 132, the majority of users (56\%) have between 50 and 200 friends, however, some users have as many as 4,476 online friends.}.
\label{fig1}
\end{figure}

\section*{Methods}
\subsection*{Survey data}
We used data from an ongoing Russian Longitudinal Panel Study of Educational and Occupational Trajectories (TrEC) \cite{malik2019russian} that tracks 4,399 students from 42 Russian regions who participated in the Programme for International Student Assessment (PISA) \cite{organisation2014pisa} in 2012. The initial TrEC sample was nationally representative for one age cohort (14--15 years old in 2012). 

In the 2018 wave, the eight-item Patient Health Questionnaire depression scale (PHQ-8) \cite{kroenke2009phq} was included in the survey to measure depressive symptoms of participants. PHQ-8 asks individuals to self-rate the frequency of various depressive symptoms over the past 2 weeks using a 4-point verbal scale: “not at all,” “several days,” “more than half the days,” and “nearly every day.” Depression symptoms are scored as the sum of all items, ranging from 0 to 24. The depression questionnaire was filled by $2,554$ participants.

The PHQ-8 scale has been shown to be a valid tool in detecting depression across various cultures \cite{richardson2010evaluation,ganguly2013patient,tsai2014patient,fatiregun2014prevalence,burdzovic2017depressive} in both clinical and population-based studies.

\subsection*{Data on online friendship}
In addition to survey data, the data set includes information on the online friendship of respondents on VK. VK was created in 2006 as a clone of Facebook and became the most popular Russian social networking site with more than $100,000$ active users. It is particularly popular among young adults: more than 90\% of 18–24 years old use it regularly \cite{fom2016online}.

VK provides an application programming interface (API) that enables the downloading of information systematically from the site. The public API could be used for research purposes according to the VK team. 

In 2018, publicly available information from VK was collected for TrEC participants. This online data is available only for those respondents who provided consent to use their VK data for research purposes (79\%). Information on non-participant was anonymized, i.e. VK ID's of participants' friends were removed. The VK data collection procedure was approved by the Institutional Review Board.

For the purposes of this study, we have analyzed the structure of 1.5 radius ego networks, i.e. the networks that include friends of a participant on VK and friendship connections between them.

\begin{figure*}[t]
\centering
\includegraphics[width=0.8\textwidth]{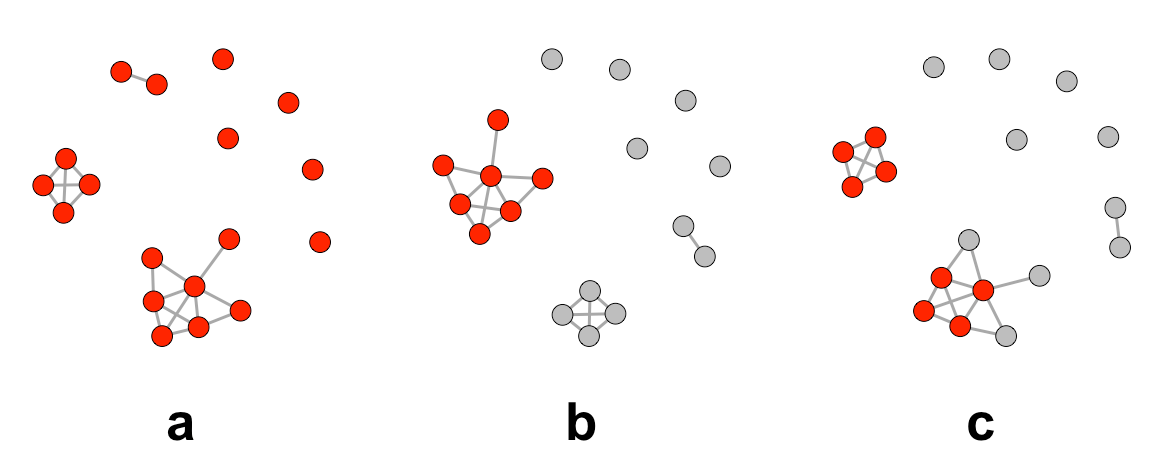} % Reduce the figure size so that it is slightly narrower than the column.
\caption{\textbf{Different approaches to identifying the core friends}. Core (red) and periphery (grey) nodes of an empirical ego network without the ego. Core friends include (a) all nodes of the network, (b) only the nodes in the largest connected component, (c) only the nodes in 3-core.}
\label{fig2}
\end{figure*}

\subsection*{Network size}
Typically to empirical social networks \cite{albert2002statistical, ugander2011anatomy}, there is a large variability in the number of users’ friends: while the majority (56\%) has between 50 and 200 friends, some users have more than 2,000 connections (see Fig. 1). Such large networks cannot represent meaningful social ties indicating that users employ different strategies when accepting or sending friend requests on social media. While for some, online friendship might indicate relatively close social connections, others probably indiscriminately accept or send requests resulting in thousands of online “friends”. To account for that, we consider separately those who have an unreasonably large number of friends and everyone else. We choose Dunbar’s number ($N = 150$) as a threshold to separate these two groups as it is thought to be a soft upper limit of the personal social network size \cite{dunbar1998social} and was empirically confirmed in both offline and online settings \cite{hill2003social, gonccalves2011modeling, wang2016critical, overgoor2020structure}. 

The exact Dunbar’s number is to a certain degree an arbitrary threshold as it is not fixed for every individual but rather is an approximate estimate. However, we find that in our particular case this choice is reasonable based on empirical observations (see Results).

\subsection*{Core}
We further assume that even for users with reasonably sized networks not all connections are necessarily equally important. One way to identify meaningful connections is to look at the cohesive structures within the ego networks. Borgatti and Everett \cite{borgatti2000models} argue that networks tend to follow the core-periphery model. It means that the network consists of two classes of nodes, namely a cohesive subgraph (the core) in which actors are connected to each other in some maximal sense and a class of actors that are loosely connected to the cohesive subgraph (the periphery). These two different structural types of social connections, in turn, might play different roles in network functioning \cite{granovetter1973strength}. Cohesive network connections are more likely to provide emotional support and resource exchange, whereas periphery ties do not have these properties as they serve different functions (e.g. binding groups together or provide information) \cite{granovetter1973strength, fingerman2009consequential}. In the case of VK, isolated friends are also more likely to be bots or accounts with fake information \cite{smirnov2016search}.

For that reason, we separate online connections into core and periphery nodes. For comparison, we use three different approaches. First, we include all friends in the core, Fig. 2a. Second, we include in the core only the nodes that are part of the largest connected component \cite{wasserman1994social}, Fig. 2b. Finally, we consider only friends that belong to the network $k$-core ($k = 3$), Fig. 2c, where $k$-core is a subgraph in which each node is adjacent to at least a minimum number, $k$, of the other nodes in this subgraph \cite{seidman1983network, wasserman1994social}. We choose $k = 3$ as it is the largest nontrivial value for our data, i.e. for $k \geq 4$ the size of $k$-core is zero for a large fraction of networks.

\section*{Results}
\subsection*{Depressive symptoms and size of the network core}
The prevalence of depression (PHQ-8 $\geq$ 10) is 16.6\% in our sample ($N = 2,554$). The average PHQ-8 score is 5.3 ($SD = 4.7$). Similar depression rates have been previously found in samples of students and young adults \cite{mikolajczyk2008prevalence, eisenberg2007prevalence, elmer2017co}. The prevalence of depression among men (51.3\% of a sample) is lower than among women: 11.5\% for men vs 21.9\% for women ($P = 2.8\cdot10^{\mathrm{-9}}$, $\chi^\mathrm{2}$-test). This is expected given that prevalence of depression among women is approximately two times larger than among men \cite{brody2018prevalence, luppa2012age}. 

The mean number of friends on VK is 190, the median is 132, similar to other online social networks \cite{ugander2011anatomy, gonccalves2011modeling}. The average size of the largest connected component is 112 nodes (the median is 70). The mean size of the network $k$-core is 117 (the median is 76). 

\begin{figure}[t]
\centering
\includegraphics[width=0.5\columnwidth]{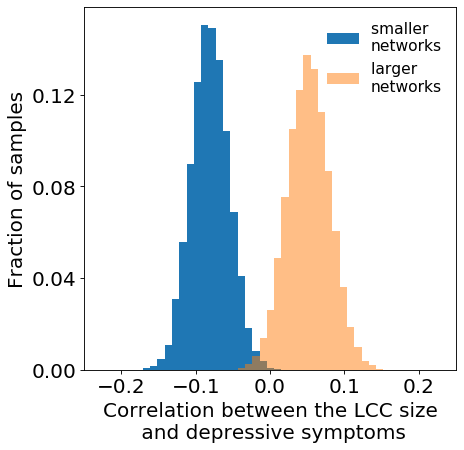} % Reduce the figure size so that it is slightly narrower than the column.
\caption{\textbf{The effect of network size on association between depressive symptoms and the size of network core}. The distribution of the correlations between network core size (largest connected component) and depressive symptoms for smaller ($N \leq 150$) and larger ego networks ($N > 150$). The distributions are from $10,000$ bootstrap simulations.}
\label{fig3}
\end{figure}

We find no association between the number of online friends and depressive symptoms for the whole sample: the correlation between PHQ-8 scores and total number of friends is Pearson’s $r = -0.01$, $P = 0.51$, for the size of the largest connected component, $r = -0.02$, $P = 0.43$, and for the core size, $r = -0.02$, $P = 0.36$.

\subsection*{Role of the network size}
We then compared the relationship between depressive symptoms and network size separately for smaller ($N \leq 150$) and larger networks ($N > 150$) using a bootstrap test. For that purpose, we repeatedly drew a sample of networks from both groups with replacement and computed Pearson correlation coefficient between the size of the largest connected component and PHQ-8 scores. The results of simulations are presented in Fig 3. The median value of correlation over 10,000 simulations for smaller networks is –0.08 and for larger networks is 0.05, the difference is significant with $P < 10^\mathrm{-3}$.

We then check for the robustness of this result with respect to changing the threshold, $N$. For that purpose we have computed the correlation between network size and depression score for networks of various sizes. We, first, computed the correlation for the smallest 30\% of all networks, i.e. networks whose size is between the 0th and 30th percentile. We then repeated the procedure by sliding the network size thresholds with a one percent step, i.e. computing correlation for networks lying between the 1st and 31st percentile, between the 2nd and 32nd, etc. The results (see Fig. 4) suggest that the relationship of interest is indeed different for smaller and larger networks. While there is a consistent significantly negative correlation for smaller networks, for larger networks it consistently does not differ from zero. Curiously, the change in the pattern seems to approximately correspond to Dunbar’s number.

For the sample of networks that are smaller or equal than 150 nodes ($N = 1,382$), we find statistically significant associations between the size of the largest connected component and depressive symptoms (Pearson’s $r = -0.08$, $P = 0.003$), and size of $k$-core and depressive symptoms ($r = -0.07$, $P = 0.006$). Intriguingly, the relationship between the overall network size and depressive symptoms remain non-significant ($r = -0.02$, $P = 0.39$). When the size of the periphery was considered separately it even correlated positively with depression: $r = 0.06$ ($P = 0.02$) for the largest connected component and $r = 0.05$ ($P = 0.05$) for the $k$-core.   

Our findings suggest that depressive symptoms of an individual are associated with their online friendship networks. The size of the network core, but not the network periphery, is negatively correlated with depressive symptoms. We also find that these results hold only for networks that do not exceed Dunbar’s number in size   and, thus, probably represent the actual social connections. 

\section*{Discussion}
Social networking sites such as Facebook, Twitter, or VK, allow their users to establish and maintain social connections with others. These connections have the potential to affect important life outcomes including an individual’s well-being and mental health. The strength and direction of these effects are still not clear. In particular, most of the studies treat online ties as homogeneous, although they might represent different kinds of relationships and as a consequence have different impacts on mental health.

In our study, we examine the relationship between the structure of online ego networks and depressive symptoms in a nationally representative sample of young adults. The results could potentially be generalized to a population level, albeit for one age cohort. We find that the size of the network is negatively associated with depressive symptoms, however, this is true only for core but not peripheral nodes. The size of the periphery is positively correlated with depression. Social comparison theory provides one possible explanation of this association. Peripheral ties may represent people we do not know personally. Previous research has shown that Facebook users with more friends who are strangers are more likely to exhibit attribution error toward those users they do not know, i.e. to attribute the positive content presented on Facebook to others’ personality, rather than situational factors \cite{chou2012they}. Thus these users are more vulnerable to social media’s positivity bias, which can lead to negative social comparison, and, in turn, emotional distress and depression \cite{feinstein2013negative, steers2014seeing, appel2016interplay, lup2015instagram}.

We also find that some users have a very large number of online friends. These online ties are unlikely to represent meaningful social connections and, perhaps not surprisingly, for users with such large networks, the size of their core network is not associated with depressive symptoms. 

These findings further support the notion that online friendship could represent different types of social relationships and, thus, online ties should be treated deferentially while investigating the association between online social integration and mental health. 

\begin{figure}[t]
\centering
\includegraphics[scale = 0.6]{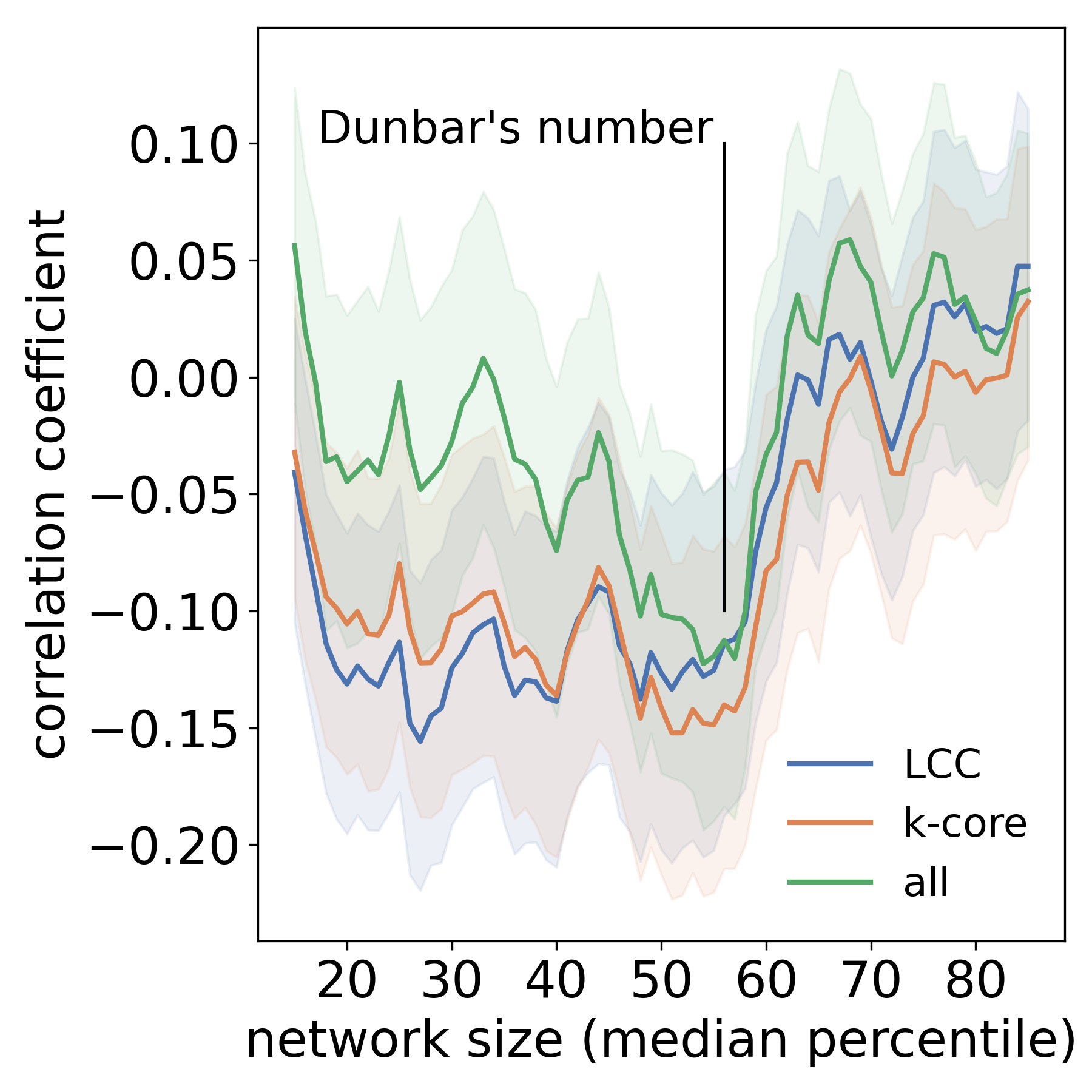} % Reduce the figure size so that it is slightly narrower than the column.
\caption{\textbf{The relationship between the network size and depressive symptoms}. The change in the strength of correlation between the number of friends and depression depending on the network size. For smaller networks, there is a consistent significantly negative correlation between the size of the network core and depressive symptoms but not for the size of the whole network. For larger networks, the correlation consistently does not differ from zero. The 90\% confidence intervals are computed via bootstrap.}
\label{fig4}
\end{figure}

\subsection*{Limitations and further research}
Our study is limited by the cross-sectional nature of the data and does not allow exploring the causal relationships between the structure of online ego networks and depressive symptoms. Furthermore, the online friendship networks are relatively stable: once added friends are rarely removed while the manifestation of depressive symptoms is typically limited in time. This might explain the weakness of found relationships. The small effect sizes should be taken into account when considering future research as it does not allow detecting potentially differential effects for different groups, e.g. depending on users’ gender, ethnicity, or socio-economic status. One potential solution is to collect data on specific groups of users, for example, those who might benefit most from online integration. 

Further research might also go beyond the information on friendship networks and consider other types of communication information, such as comments, direct messages, or likes. This might serve as a better proxy for social integration, see, for example, \cite{hobbs2016online}.

Overall, our results suggest that future studies of the effects of online social integration on depression should focus on refined measurements of social integration to identify \textit{actual} and \textit{active} connections at several time points. 

\section*{Acknowledgements}
This work was supported by a grant from the Russian Science Foundation (project №19-18-00271).

The data of the Russian panel study “Trajectories in Education and Career” (TrEC http://trec.hse.ru/) is presented in this work. The TrEC project is supported by the Basic Research Programme of the National Research University Higher School of Economics.
\bigskip
\bibliographystyle{unsrt} 
\bibliography{paper}

\end{document}